\documentclass[aps,preprint,preprintnumbers,amsmath,amssymb,amsfonts,groupedaddress,superscriptaddress,showpacs,showkeys,floatfix]{revtex4-1}

\usepackage{graphicx}
\usepackage{subfigure}
\usepackage{bm}
\usepackage{amsmath}
\usepackage{color}
\usepackage{array}

\newcommand{\bmath}{\begin{mathletters}}
\newcommand{\emath}{\end{mathletters}}
\newcommand{\be}{\begin{eqnarray}}
\newcommand{\ee}{\end{eqnarray}}
\newcommand{\ba}{\begin{array}}
\newcommand{\ea}{\end{array}}
\newcommand{\la}{\langle}

\newcommand{\ra}{\rangle}

\newcommand{\h}{\hbar}
\newcommand{\pr}{\prime}

\newcommand{\calD} {\mathcal D}

\newcommand{\rmd} {\mathrm{d}}

\begin{document}
\title{The experimental realization of high-fidelity `shortcut-to-adiabaticity' quantum gates in a superconducting Xmon qubit}

\author{Tenghui Wang}
 \affiliation{Physics Department, Zhejiang University, Hangzhou, 310027, China}
 \author{Zhenxing Zhang}
 \affiliation{Physics Department, Zhejiang University, Hangzhou, 310027, China}
 \author{Liang Xiang}
 \affiliation{Physics Department, Zhejiang University, Hangzhou, 310027, China}
 \author{Zhilong Jia}
 \affiliation{Key Laboratory of Quantum Information, University of Science and Technology of China, Hefei, 230026, China}
 \author{Peng Duan}
 \affiliation{Key Laboratory of Quantum Information, University of Science and Technology of China, Hefei, 230026, China}
 \author{Weizhou Cai}
 \affiliation{Center for Quantum Information, Institute for Interdisciplinary Information Sciences, Tsinghua University, Beijing, 100084, China}
 \author{Zhihao Gong}
 \affiliation{Physics Department, Zhejiang University, Hangzhou, 310027, China}
 \author{Zhiwen Zong}
 \affiliation{Physics Department, Zhejiang University, Hangzhou, 310027, China}
 \author{Mengmeng Wu}
 \affiliation{Physics Department, Zhejiang University, Hangzhou, 310027, China}
 \author{Jianlan Wu }
 \email{jianlanwu@zju.edu.cn}
 \affiliation{Physics Department, Zhejiang University, Hangzhou, 310027, China}
 \author{Luyan Sun}
 \affiliation{Center for Quantum Information, Institute for Interdisciplinary Information Sciences, Tsinghua University, Beijing, 100084, China}
 \author{Yi Yin}
 \email{yiyin@zju.edu.cn}
 \affiliation{Physics Department, Zhejiang University, Hangzhou, 310027, China}
 \author{Guoping Guo}
 \email{gpguo@ustc.edu.cn}
 \affiliation{Key Laboratory of Quantum Information, University of Science and Technology of China, Hefei, 230026, China}

\begin{abstract}

Based on a `shortcut-to-adiabaticity' (STA) scheme, we theoretically design and experimentally realize a set of high-fidelity single-qubit quantum gates
in a superconducting Xmon qubit system. Through a precise microwave control, the qubit is driven to follow a fast `adiabatic'
trajectory with the assistance of a counter-diabatic field and the correction of derivative removal by adiabatic gates.
The experimental measurements of quantum process tomography and interleaved randomized benchmarking show that the process fidelities of our
STA quantum gates are higher than $94.9\%$ and the gate fidelities are higher than $99.8\%$, very close to the state-of-art gate fidelity
of $99.9\%$. An alternate of high-fidelity quantum gates is successfully achieved under the STA protocol.

\end{abstract}

\maketitle

\section{Introduction}
\label{sec1}

Quantum computation and quantum information processing are programmed through sequential operations of various quantum gates,
which are built bottom up from simple but fundamental single- and two-qubit gates~\cite{ChuangBook,LaddNat10}.
A gate error has to be controlled below a fault-tolerant threshold in scale-up quantum computation. 
Since this error threshold is usually small ($0.1\%\sim 1\%$), the experimental realization of
high fidelity quantum gates is an essential task in various artificial quantum systems
such as nuclear magnetic resonance~\cite{RyanNJP09,LuNPJQI17}, ion traps~\cite{BenhelmNatPhys08} and superconducting circuits~\cite{BarendsNat14}.

A unitary transformation occurs when a single- or multi-qubit system is operated by a quantum gate. For a single
qubit, such a unitary transformation can be viewed as a rotation of a qubit vector, which can be mapped onto a spin,
on the Bloch sphere. Subject to an external magnetic field along a fixed direction, the rotation angle of the spin
is controlled by adjusting the amplitude of the magnetic field over time.
By mapping a driving pulse, e.g., Gaussian-shaped, onto a magnetic field,
we can build a single-qubit quantum gate based on the above
scheme. This standard approach has been applied in almost all the artificial quantum devices.
In superconducting qubit systems, the highest single-qubit fidelity is achieved at the level of $>99.9\%$
by optimizing the pulse amplitude and frequency~\cite{BarendsNat14}.

An alternative way of constructing quantum gates is to change the direction of the magnetic field over time.
In a special moving reference frame, the motion of the spin can be highly simplified.
In a quantum adiabatic operation, the qubit is kept at its instantaneous eigenstates.
With respect to the instantaneous eigen basis, the qubit vector is rotated along a fixed latitude
on a moving Bloch sphere by accumulating dynamic and geometric phases~\cite{SjoqvistPhy08,ZhuPRA05}.
At the end of such an quantum adiabatic operation, an arbitrary quantum gate is realized by
the combined effect of a simple spin rotation in the moving frame and the rotation of the reference frame.

However, an ideal adiabatic operation can only be performed with an infinitely slow speed.
A practically adiabatic implementation inevitably includes errors due to non-adiabatic transition
and quantum dissipation. The associated long operation time leads to a technical difficulty
in scale-up quantum computation. The shortcut-to-adiabaticity (STA) procedure has been proposed
to solve these problems by introducing a counter-diabatic field in addition to the reference
fast `adiabatic' field~\cite{DemirplakJPCA03,BerryJPhysA09,XChenPRL2010,MasudaPRS10,Torrontegui13AAMOPhy,CampoPRL12,CampoPRL13,TongSR15,SantosSciRep15}.
The qubit system is driven to follow the reference `adiabatic' trajectory
by suppressing the non-adiabatic transitions. As the quantum operation is accelerated ten to
hundred times, the decoherence induced error can be significantly reduced. The STA protocol
has been well implemented experimentally soon after it was proposed
theoretically~\cite{BasonNatPhy11, JFZhangPRL13,AnNatCommu16,ZhouNatPhys16,DuYXNatCommu16}.
In our recent experiments with a superconducting phase qubit, we successfully measured the Berry phase~\cite{ZZXPRA17}
and achieved a high-fidelity state transfer under the STA protocol~\cite{SCPMA2018}. The state transfer technique
was further applied to simulate a quantum topological phase transition~\cite{SCPMA2018}.

In this paper, we extend our previous work of quantum state transfer for the purpose of single-qubit
STA quantum gates. Our theoretical design shares the same principle as in a recent proposal in
the system of NV centers~\cite{LiangPRA16}. The detailed driving pulse is different 
but preserves the utilization of the phase accumulation in the fast `adiabatic' evolution. 
With the improvement from a superconducting phase to Xmon qubit, the high-fidelity STA quantum gate is successfully
achieved, as demonstrated by our quantum process tomography and interleaved randomized
benchmarking measurements. For our examples of the rotations about $X$-, $Y$- and $Z$-axes
and the Hadamard gate, the gate fidelity is consistently higher than 99.8\%, which promises
an alternative choice of quantum gates for a practical application.

\section{Theory}
\label{sec2}

In this section, we demonstrate our theoretical design of a general single-qubit gate 
performed under the `shortcut-to-adiabaticity' (STA) protocol.

\subsection{Adiabatic  Quantum  Gate}
\label{sec2a}

A single qubit of $\{|0\rangle, |1\rangle\}$ can be mapped onto a spin-$1/2$ particle $\{|\uparrow\rangle, |\downarrow\rangle\}$
driven by an external field~\cite{ChuangBook}. In the rotating frame, the time-dependent Hamiltonian is written as
\be
 H_0(t)=\hbar \bm B_0(t)\cdot\bm \sigma/2,
\label{eq_01}
\ee
where ${\boldsymbol B}_0(t) = \Omega(t)(\sin\theta(t)\cos\phi(t),\sin\theta(t)\sin\phi(t),\cos\theta(t))$
is the vector of an external field and  ${\boldsymbol \sigma} = (\sigma_x, \sigma_y, \sigma_z)$ is the vector of Pauli matrices.
The amplitude $\Omega(t)$, the polar angle $\theta(t)$ and the azimuthal angle
$\phi(t)$ are modulated by microwave pulse sequences in our experiment~\cite{ZZXPRA17,SCPMA2018}.
At a given time $t$, the instantaneous eigenstates,
$\{|\psi_+(t)\rangle, |\psi_-(t)\rangle\}$, are obtained by a rotation of the reference states, $\{|\uparrow\rangle, |\downarrow\rangle\}$,
where the rotation matrix to change the frame is given by
\be
S(t) =
\left(\ba{cc} \cos\frac{\theta(t)}{2} & \sin\frac{\theta(t)}{2} e^{-i\phi(t)} \\
 - \sin\frac{\theta(t)}{2}e^{i\phi(t)} & \cos\frac{\theta(t)}{2}
 \ea \right).
\label{eq_02}
\ee
For an extremely slow variation of the external field, the spin-$1/2$ particle remains at the same instantaneous eigenstate, $|\psi_{+/-}(t)\rangle$,
if it is prepared at $|\psi_{+/-}(0)\rangle$ initially.
During this adiabatic propagation, only the dynamic and geometric phases are accumulated. With respect to the instantaneous eigen basis,
a unitary transformation is thus defined as $U_{\mathrm{ad}}(t)=|\psi_+(t)\rangle U_{\mathrm{ad}; ++}(t)\langle\psi_+(0)|
+|\psi_-(t)\rangle U_{\mathrm{ad}; --}(t)\langle\psi_-(0)|$. In a matrix representation, this adiabatic unitary transformation is explicitly written as
\be
U_{\mathrm {ad}}(t) =
\left(\ba{cc} e^{i\varphi_d(t)+i\gamma_{+}(t)} & 0 \\
0 & e^{-i\varphi_d(t)+i\gamma_{-}(t)}
 \ea \right),
\label{eq_03}
\ee
where $\varphi_d(t)= -(1/2)\int_0^t \Omega(\tau)d\tau $ and $\gamma_{\pm}(t) =i\int_0^t\langle\psi_{\pm}(\tau)|\partial_{\tau}|\psi_{\pm}(\tau)\rangle d\tau$
are the dynamic and geometric phases, respectively. Here we consider a special form of the amplitude evolution,
\be
\Omega(t) = A\sin\left(\frac{2\pi t}{T}\right),
\label{eq_04}
\ee
where the parameter $T$ is the time of our quantum operation. The accumulated dynamic phases vanish, i.e., $\varphi_d(T)=0$.
After a global phase shift, the  unitary transformation is simplified to
\be
U_{\mathrm{ad}}(T) =
\left(\ba{cc} 1 & 0 \\
0 & e^{-i\Delta\gamma(T)}
 \ea \right),
\label{eq_05}
\ee
with $\Delta\gamma(T) = \gamma_{+}(T) - \gamma_{-}(T)$. If the initial preparation and final measurement are
performed in the reference basis of $\{|\uparrow\rangle, |\downarrow\rangle\}$, the combined unitary transformation is given by
\be
U = S^+(T)U_{\mathrm {ad}}(T)S(0),
\label{eq_06}
\ee
which leads to an arbitrary single-qubit quantum gate~\cite{ChuangBook}. This adiabatic construction can be straightforwardly extended to multi-qubit gates,
which will be studied in the future.

\subsection{STA Protocol}
\label{sec2b}

In practice, the remaining non-adiabatic transition introduces an inevitable error for an adiabatic quantum gate. In the STA protocol, an additional
counter-diabatic Hamiltonian is applied to cancel this non-adiabatic error~\cite{DemirplakJPCA03,BerryJPhysA09,XChenPRL2010,MasudaPRS10,Torrontegui13AAMOPhy,CampoPRL12,CampoPRL13,TongSR15,SantosSciRep15}.
A general time-dependent Hamiltonian $H_0(t)$ can be expanded in its instantaneous eigen basis,
giving $H_0(t)=\sum_n\epsilon_n(t) |\psi_n(t)\rangle\langle \psi_n(t)|$ with $\epsilon_n(t)$ the $n$-th eigenenergy and  $|\psi_n(t)\rangle$ the $n$-th eigenstate.
Accordingly, the counter-diabatic Hamiltonian $H_{\mathrm{cd}}(t)$ is formally written as~\cite{BerryJPhysA09}
\be
H_{\mathrm{cd}}(t) = i\hbar\sum_n \big[|\partial_t \psi_n(t)\rangle\langle \psi_n(t)|-\langle \psi_n(t)| \partial_t \psi_n(t)\rangle | \psi_n(t) \rangle \langle \psi_n(t)| \big],
\label{eq_07}
\ee
which suppresses the non-adiabatic transition for each eigenstate $|\psi_n(t)\rangle$.
The quantum system driven $H(t)=H_0(t)+H_{\mathrm{cd}}(t)$ rigorously evolves along the
instantaneous eigenstates of $H_0(t)$. The time propagator becomes exactly diagonal in the instantaneous eigen basis, i.e.,
\be
U_\mathrm{STA}(t) = \sum_n |\psi_n(t)\rangle U_{\mathrm{STA}; nn}(t) \langle \psi_n(0)|
\label{eq_07a}
\ee
The adiabatic quantum gate introduced in Eq.~(\ref{eq_06}) is thus changed to a STA  quantum gate,
\be
U = S^+(T)U_{\mathrm {STA}}(T)S(0),
\label{eq_07b}
\ee
by replacing $U_\mathrm{ad}(T)$ with $U_\mathrm{STA}(T)$.
As the quantum operation time $T$ is decreased, the error induced by relaxation and decoherence can be significantly reduced while
the non-adiabatic error is fully suppressed in the ideal scenario. The STA protocol provides an alternative design of
high-fidelity quantum gates~\cite{LiangPRA16}.

For the spin-$1/2$ particle under the Hamiltonian in Eq.~(\ref{eq_01}), the counter-diabatic Hamitlonian follows
a similar form,
\be
H_\mathrm{cd}(t)=\hbar \bm B_\mathrm{cd}(t)\cdot\bm \sigma/2.
\label{eq_08}
\ee
Through a tedious but straightforward derivation from Eq.~(\ref{eq_07}),
the three elements of the counter-diabatic field $\bm B_\mathrm{cd}(t)$ are explicitly given by
\be
\left\{ \ba{ccl} B_{\mathrm{cd}; x}(t) &=&   -\dot{\theta}(t)\sin\phi(t)-\dot{\phi}(t)\sin\theta(t)\cos\theta(t)\cos\phi(t) \\
B_{\mathrm{cd}; y}(t) &=&  \dot{\theta}(t)\cos\phi(t)-\dot{\phi}(t)\sin\theta(t)\cos\theta(t)\sin\phi(t) \\
B_{\mathrm{cd}; z}(t) &=&  \dot{\phi}(t)\sin^2\theta(t) \ea\right. .
\label{eq_09}
\ee
Equation~(\ref{eq_09}) can be further organized into a cross product form as~\cite{BerryJPhysA09,ZZXPRA17,SCPMA2018}
\be
\bm B_\mathrm{cd}(t) = \frac{1}{|\bm B_0(t)|^2}\bm B_0(t)\times\dot{\bm B}_0(t),
\label{eq_10}
\ee
which is always orthogonal to the reference field $\bm B_0(t)$. By applying the external field,
$\boldsymbol B(t)=\boldsymbol B_0(t)+\boldsymbol B_\mathrm{cd}(t)$, to a single qubit,
the STA gates will be testified experimentally in our Xmon qubit system.

\subsection{DRAG Correction}
\label{sec2c}

In many artificial systems, the influence of higher excited states cannot be fully ignored so that
the two-level qubit has to be re-modelled as a multi-level anharmonic oscillator~\cite{ChuangBook,BarendsPRL13}.
For example, the Hamiltonian of a three-level anharmonic oscillator in the rotating frame is written as~\cite{SCPMA2018}
\be
H(t) =  \frac{\h}{2} \bm{B}(t)\cdot\bm{S}+\h\Delta_2|2\ra\la 2|,
\label{eq_11}
\ee
where the operator vector $\boldsymbol S$ is given by
\be
\left\{\ba{ccl}
S_x &=& \sum_{n=0}^{1}\sqrt{n+1}\left(|n+1\rangle\langle n|+|n\rangle\langle n+1|\right) \\
S_y &=& \sum_{n=0}^{1}\sqrt{n+1}\left(i|n+1\rangle\langle n|-i|n\rangle\langle n+1|\right)  \\
S_z &=& \sum_{n=0}^2 (1-2n) |n\rangle\langle n| \ea \right.
\label{eq_12}
\ee
and $\Delta_2$ is an anharmonic parameter. In the STA protocol, the external field is given by
$\bm B(t) = \bm B_0(t) +\bm B_\mathrm{cd}(t)$. A technical treatment
is to apply the derivative removal by adiabatic gates (DRAG) method, which decouples the interaction between
the lowest two levels (qubit) and higher excited states~\cite{SCPMA2018,MotzoiPRL09,GambettaPRA11,LuceroPRA10,ChowPRA10}.
With the increment of another field,
$\bm B_{\rm d}(t)=(B_{\rmd; x}(t), B_{\rmd; y}(t), B_{\rmd; z}(t))$, the total external field is
changed to $\bm B^\pr(t)=\bm B(t)+\bm B_{\rm d}(t)$ and the total Hamiltonian in Eq.~(\ref{eq_11})
is modified to be $H^\pr(t)= (\h/2)\bm B^\pr(t)\cdot \bm S+\hbar \Delta_2 |2\rangle\langle 2|$.
In addition, we introduce the DRAG frame ($\calD$-frame), in which the total Hamiltonian is transformed into
\be
H_\calD(t)=\calD^+(t)H^\pr(t)\calD(t)+i\dot{\calD}^+(t)\calD(t).
\label{eq_13}
\ee
where $\calD(t)$ is a unitary operator. The density matrix in the $\calD$-frame is given by $\rho_\calD(t)=\calD(t)\rho(t)\calD^+(t)$.
With a delicate design of $\bm B^\pr(t)$ and $\calD(t)$, the transformed Hamiltonian $H_\calD(t)$  is factorized into
\be
H_\calD(t) =\left[\varepsilon(t)+ \frac{\hbar}{2}\bm B(t)\cdot \bm \sigma\right] \oplus \varepsilon_2(t)|2\rangle\langle 2|,
\label{eq_14}
\ee
where $\varepsilon(t)$ and $\varepsilon_2(t)$ are two shifted energies. The qubit subspace of $\{|0\rangle, |1\rangle\}$
is decoupled with the second excited state $|2\rangle$. To avoid an artifact of the $\calD$-frame, we would expect
an requirement of
\be
\calD(t=0)=1~~~\mathrm{and}~~~\calD(t=T)=1,
\label{eq_15}
\ee
so that the density matrices at the initial and final moments of the quantum operation are unaffected, i.e.,
$\rho_\calD(0)=\rho(0)$ and $\rho_\calD(T)=\rho(T)$. In the DRAG method, $\bm B^\pr(t)$ and $\calD(t)$ are evaluated
by a perturbation approach with the assumption of a large anharmoncity, i.e., $|\Delta_2| \gg |\bm B(t)|$. On the first order
correction, the DRAG field $\bm B_\mathrm{d}(t)$ is explicitly given by~\cite{GambettaPRA11,SCPMA2018}
\be
\left\{\ba{ccl} B_{\rmd; x}(t) &=& \frac{1}{2\Delta_2}\left[\dot{B}_{y}(t)-B_{z}(t)B_{x}(t)\right] \\
B_{\rmd; y}(t) &=& -\frac{1}{2\Delta_2}\left[\dot{B}_{x}(t)+B_{z}(t)B_{y}(t)\right]  \\
B_{\rmd; z}(t) &=& 0 \ea \right. ,
\label{eq_16}
\ee
under a presumption of $B_{\rmd; z}(t) =0$. In our experiment, the Xmon qubit is driven by the total
external field, $\bm B_\mathrm{tot}(t) = \bm B_0(t) +\bm B_\mathrm{cd}(t)+\bm B_\mathrm{d}(t)$, under the STA
protocol and with the DRAG correction.

\begin{figure}[tp]
\centering
\includegraphics[width=0.55\columnwidth]{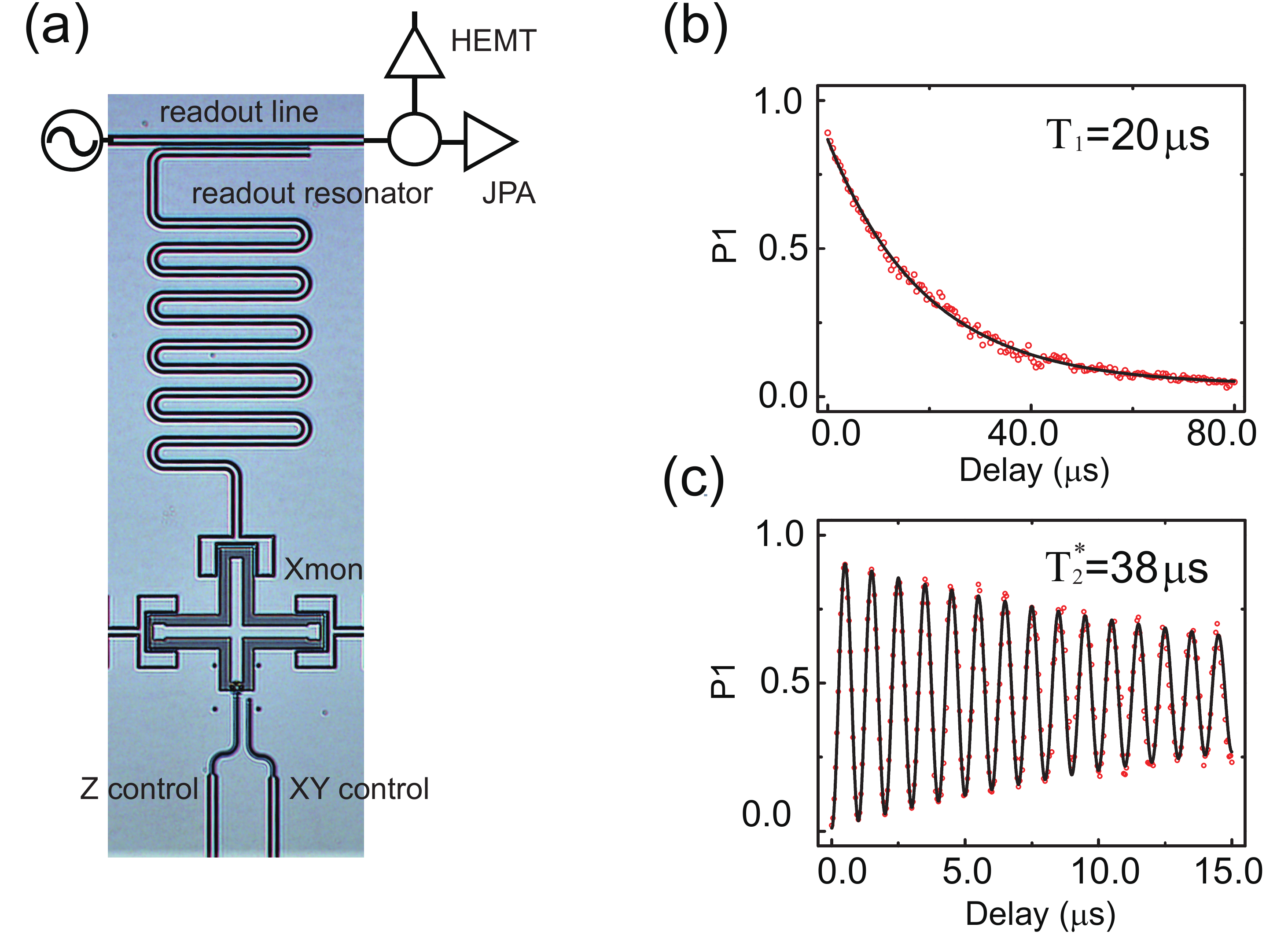}
\caption{(a) An optical micrograph of a single cross-shaped Xmon qubit. 
(b) Energy decay of the qubit, giving a relaxation time of $T_1=20$ $\mu$s. (c) Ramsey fringes of the qubit, giving a pure decoherence time of $T_2^{*}=38$ $\mu$s.
 }
\label{fig_n01}
\end{figure}

\section{Experimental Setup}
\label{sec3}

A cross shaped transmon (or called Xmon) qubit~\cite{BarendsPRL13,BarendsNat14,KellyNat15} is applied in this experiment.
The Xmon qubit sample is fabricated on a silicon substrate. After initially cleaned
in buffered hydrofluoric acid to remove the native oxide, the substrate is immediately loaded into a high vacuum
electron beam evaporator, followed by a deposition of an aluminum (Al) film. The
superconducting resonators and control lines are patterned using photolithography in a wafer stepper
and etched with BCl$_3$/Cl$_2$ in an inductively coupled plasma (ICP) dry etcher.
The superconducting Josephson junctions are patterned with an electron beam lithography and
developed with Al double-angle evaporation. An additional `bandage' DC electrical
contact is fabricated to reduce the capacitive loss~\cite{DunsworthAPL17}.

Figure~\ref{fig_n01}(a) displays an optical micrograph of a single Xmon qubit. Four arms
of the cross are connected to different elements for separate functions
of coupling, control and readout. At the bottom of the cross, a flux current ($Z$ control) line
biases the qubit at a resonance frequency of $\omega_{10}/2\pi=4.85$ GHz, which is the
energy difference between the ground  ($|{0}\rangle$) and excited  ($|{1}\rangle$) states
of the qubit. The qubit nonlinearlity is $\Delta_2/2\pi=-253$ MHz. Another $XY$ control line
provides a microwave drive signal to the qubit to
manipulate the qubit state~\cite{BarendsPRL13,BarendsNat14,KellyNat15}. The top arm of the cross is coupled to a readout resonator
whose bare frequency is $\omega_\mathrm{r}/2\pi=6.56$ GHz. By sending a microwave signal through
the readout line, we can detect the qubit state information from the dispersive interaction
between the qubit and readout resonator. The readout signal is followed by a Josephson parametric amplifier (JPA)~\cite{RoyAPL15,XiaoYuanPRL16}
and a high electron mobility transistor (HEMT) for a high fidelity measurement. By heralding the ground state~\cite{JohnsonPRL12}, the
readout fidelity for the ground state $|0\rangle$ and excited state $|1\rangle$ are $99.8\%$ and $95.1\%$,
respectively. With the qubit biased at a sweet point here, the coherence is characterized
by a relaxation time, $T_1=20$ $\mu$s, and a pure decoherence time, $T_2^{*}=38$ $\mu$s (see Figs.~\ref{fig_n01}(b) and~\ref{fig_n01}(c)).
Our current sample is designed as a linear array with six qubits. All the qubtis have comparable
values of $T_1$ and $T_2^*$. The qubit chip is mounted in a sample box
and cooled in a dilution refrigerator whose base temperature is $\sim 10$ mK.

\begin{figure}[tp]
\includegraphics[width=0.6\columnwidth]{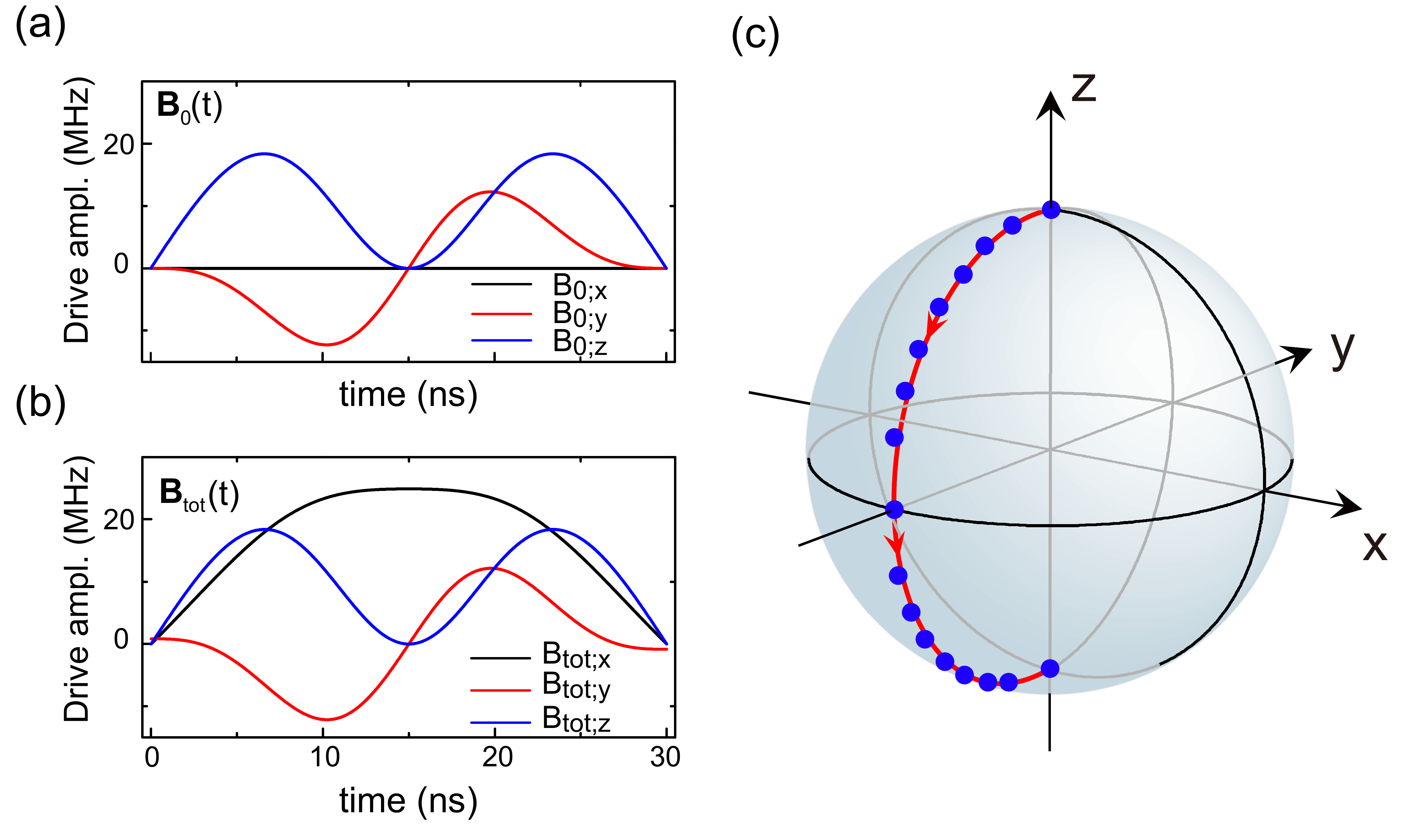}
\caption{(a) The reference `adiabatic' and (b) total (with the counter-diabatic and DRAG corrections) fields for a $\pi$ rotation about the $X$-axis.
The maximum drive amplitude is $A/2\pi = 20$ MHz and the operation time is $T = 30$ ns.
(c) The fast `adiabatic' trajectory of the qubit vector for the initial state at $|0\rangle$.
The ideal result is shown in a red arrowed curve on the Bloch sphere while the experimental result after the correction of the measurement error
is shown in blue dots.
 }
\label{fig_n02}
\end{figure}

\section{Results}
\label{sec4}

In this section, we present our experimental realization of various single-qubit STA quantum gates.

\subsection{$X_\pi$ and $X_{\pi/2}$ Rotations}
\label{sec4a}

The unitary matrices representing the $\pi$ and $\pi/2$ rotations about the $X$-axis ($X_\pi$ and $X_{\pi/2}$ rotations) are explicitly written as~\cite{ChuangBook}
\be
U_{X_\pi}=\left(\ba{cc} 0 & -i\\
                  -i & 0 \ea\right)~~~\mathrm{and}~~~
U_{X_{\pi/2}}=\frac{\sqrt{2}}{2}\left(\ba{cc} 1& -i\\
                                       -i&  1\ea\right).
\label{eq_17}
\ee
To design the $X_\pi$ rotation, the reference `adiabatic' field $\bm B_0(t)$ is specified as
\be
\left\{\ba{ccl} B_{0; x}(t) &=& 0  \\
B_{0; y}(t) &=& -\Omega(t)\sin\theta(t)\\
B_{0; z}(t) &=& \Omega(t) \cos\theta(t)\\
 \ea \right. .
\label{eq_18}
\ee
The drive amplitude, polar and azimuthal angles are $\Omega(t) = A \sin(2\pi t/T)$,
$\theta(t)= (\pi/2) [1-\cos(\pi t/T)]$, and $\phi(t)=-\pi/2$, respectively. In our experiment,
we set the pulse length (operation time) at $T = 30$ ns and the maximum drive amplitude at $A/2\pi = 20$ MHz.
The same two parameters will be used in other STA gates.
The pulse length is comparable to the typical value of a truncated Gaussian pulse.
In principle, these two parameters can be modified independently under the STA protocol.
The counter-diabatic field $\bm B_\mathrm{cd}(t)$ and the DRAG field $\bm B_\mathrm{d}(t)$ are
calculated using Eqs.~(\ref{eq_09}) and~(\ref{eq_16}). Due to the limitation of space, we will not
present the analytical forms of $\bm B_\mathrm{cd}(t)$ and $\bm B_\mathrm{d}(t)$.
In Figs.~\ref{fig_n02}(a) and~\ref{fig_n02}(b), we plot the $x$-, $y$- and $z$-components of
the reference field $\bm B_0(t)$ and the total field $\bm B_\mathrm{tot}(t)=\bm B_0(t)+\bm B_\mathrm{cd}(t)+\bm B_\mathrm{d}(t)$.
As a comparison, the major difference between the two fields appears in their $x$-components.
With the condition of $|A|\ll |\Delta_2|$, the DRAG correction is a minor effect.
For an initial preparation at the spin-up state ($|\uparrow\rangle=|0\rangle$),
the fast `adiabatic' trajectory of the qubit is shown in Fig.~\ref{fig_n02}(c).
In an ideal scenario, the qubit vector evolves from the north to south pole along $270^{\circ}$-longitude of the Bloch sphere,
and the final qubit state is the spin-down state ($|\downarrow\rangle=|1\rangle$).
Figure~\ref{fig_n02}(c) shows that this trajectory can be excellently generated under the STA control
field $\bm B_\mathrm{tot}(t)$~\cite{SCPMA2018}.

\begin{figure}[tp]
\includegraphics[width=0.6\columnwidth]{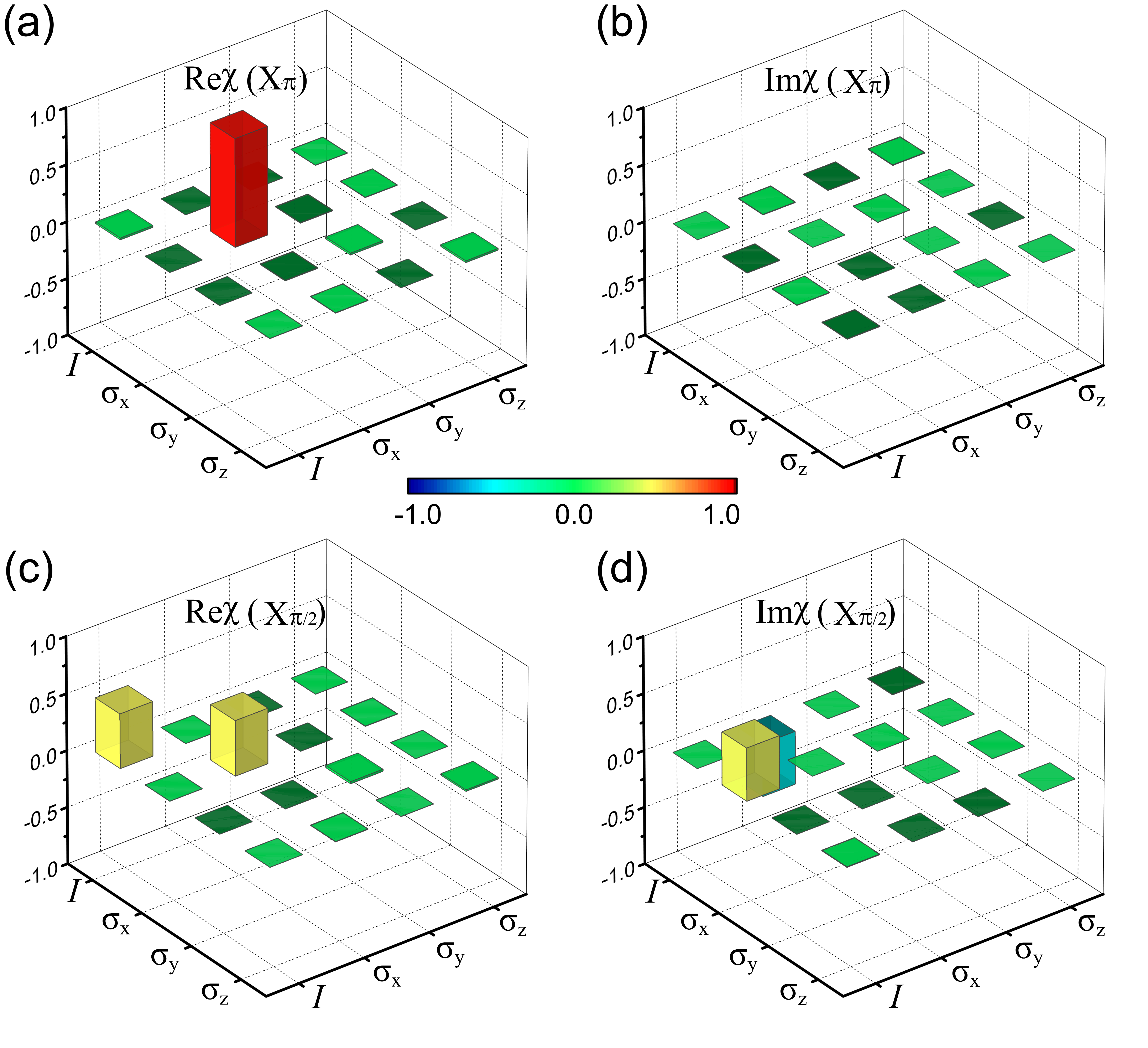}
\caption{The experimental measurement of $\chi$ matrices for (a-b) $X_\pi$ and (c-d) $X_{\pi/2}$ rotations.
The left and right panels are the real and imaginary parts of the two $\chi$ matrices, respectively. }
\label{fig_n03}
\end{figure}

With the consideration of the errors in state preparation, STA operation and readout, the output state is obtained
through a map of the input state~\cite{ChuangBook}, i.e.,
\be
\text{\large $\varepsilon$}: \rho\mapsto \text{\large $\varepsilon$}(\rho) = \sum_{i=1}^4 E_i\rho E_i^+,
\label{eq_19}
\ee
with $\rho$ the initial density matrix of the qubit. Each linear operators $E_{i=1, \cdots, 4}$ can be
expanded over a fixed set of operators, $\{\tilde{E}_m=I,\sigma_x, \sigma_y, \sigma_z\}$,
giving $E_i = \sum_m e_{im} \tilde{E}_m$. The output density matrix is rewritten as
\be
\text{\large $\varepsilon$}(\rho) = \sum_{mn} \chi_{mn} \tilde{E}_m\rho \tilde{E}_n^+
\label{eq_20}
\ee
with $\chi_{mn} = \sum_i e_{im}e_{in}^\ast$. The $\chi$ matrix thus completely characterizes the behavior of a specific gate.
To experimentally determine the $\chi$ matrix, we perform the quantum process tomography (QPT) by selecting
6 different initial states, $\{|0\rangle, |1\rangle, (|0\rangle \pm |1\rangle )/\sqrt{2},
(|0\rangle \pm i|1\rangle )/\sqrt{2}\}$~\cite{BialczakNatPhys10,YamamotoPRB10,ChuangBook}. Each input state is driven by $\bm B_\mathrm{tot}(t)$ and
the output state is measured by the quantum state tomography (QST) method. The $\chi$ matrix is then numerically calculated by
solving Eq.~(\ref{eq_20}). For the STA $X_\pi$-gate,  the experimental result of the $\chi(X_\pi)$ matrix is plotted
in Figs.~\ref{fig_n03}(a) and~\ref{fig_n03}(b). Consistent with the theoretical prediction of an ideal
$X_\pi$-gate, the dominant element of the $\chi(X_\pi)$ matrix is the operator of $\sigma_x$.
To quantify the fidelity of the whole quantum process, we calculate the process fidelity using~\cite{ChuangBook}
\be
F_P = \mathrm{Tr}\{\chi \chi_\mathrm{ideal}\}.
\label{eq_21}
\ee
The experimental result is $F_P(X_\pi) = 95.21\%$. To exclude the errors in state preparation and readout,
we perform an interleaved randomized benchmarking measurement (see Sec.~\ref{sec4d}), which gives the gate fidelity
of the STA $X_\pi$ rotation at $F_g(X_\pi)= 99.82\%$. This number is very close to the current highest fidelity
of a Xmon qubit~\cite{BarendsNat14}, and the 0.1\% deviation could be improved by the future optimization of our system.

To design the $X_{\pi/2}$ rotation, we take the same reference `adiabatic' field $\bm B_0(t)$ except for
that the azimuthal angle is changed to $\theta(t)= (\pi/4) [1-\cos(\pi t/T)]$. The counter-diabatic
and DRAG fields, $\bm B_\mathrm{cd}(t)$ and $\bm B_\mathrm{d}(t)$, are analytically calculated accordingly.
After the QPT measurement, the experimentally reconstructed $\chi(X_{\pi/2})$ matrix is
plotted in Figs.~\ref{fig_n03}(c) and~\ref{fig_n03}(d), agreeing excellently with the theoretical prediction
of an ideal $X_{\pi/2}$ gate. As compared to the $\chi(X_{\pi})$ matrix, the $\chi(X_{\pi/2})$ matrix
includes auto and cross correlations between the operators of $I$ and $\sigma_x$.
The experimental measurement shows that the process and gate fidelities of our STA $X_{\pi/2}$ rotation
are $F_P(X_{\pi/2})=95.03\%$ and $F_g(X_{\pi/2})=99.81\%$.

\begin{figure}[tp]
\includegraphics[width=0.6\columnwidth]{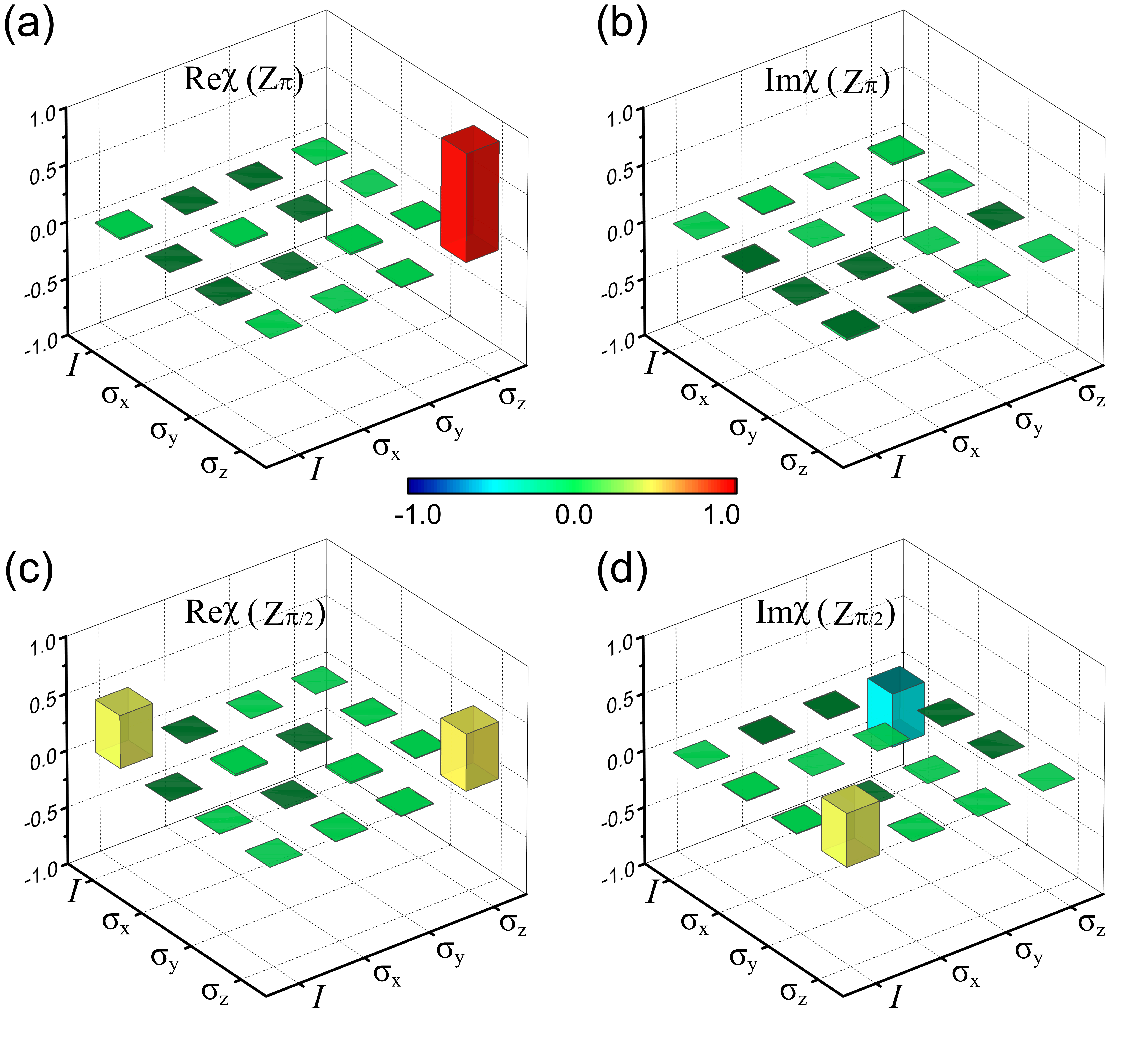}
\caption{The experimental measurement of $\chi$ matrices for (a-b) $Z_\pi$ and (c-d) $Z_{\pi/2}$ rotations.
The left and right panels are the real and imaginary parts of the two $\chi$ matrices, respectively.}
\label{fig_n04}
\end{figure}

\subsection{$Z_\pi$ and $Z_{\pi/2}$ Rotations}
\label{sec4b}

The second group of STA quantum gates we inspect are the $\pi$ and $\pi/2$ rotations about $Z$-axis.
The corresponding unitary matrices are~\cite{ChuangBook}
\be
U_{Z_\pi} = \left(\ba{cc} -i &0\\
                    0  & i \ea \right),~~~\mathrm{and}~~~
U_{Z_{\pi/2}} = \left(\ba{cc} e^{-i\pi/4}&0\\
                        0& e^{i\pi/4} \ea \right).
\label{eq_22}
\ee
To design these two gates, the reference `adiabatic' field $\bm B_0(t)$ is specified as
\be
\left\{\ba{ccl} B_{0; x}(t) &=& \Omega(t) \cos\phi(t) \\
B_{0; y}(t) &=& \Omega(t) \sin\phi(t)\\
B_{0; z}(t) &=& 0
 \ea \right.,
\label{eq_23}
\ee
where the drive amplitude is $\Omega(t) = A \sin(2\pi t/T)$ and the polar angle is $\theta(t)=\pi/2$.
The azimuthal angles for $Z_\pi$ and $Z_{\pi/2}$ rotations are $\phi(t)= (\pi/2) [1-\cos(\pi t/T)]$
and $\phi(t)= (\pi/4) [1-\cos(\pi t/T)]$, respectively. The control parameters, $A$ and $T$, are the same as
those in the $X$-rotation gates. The counter-diabatic and DRAG fields, $\bm B_\mathrm{cd}(t)$ and
$\bm B_\mathrm{d}(t)$, are also analytically calculated for the experimental generation.
The QPT measurements of $\chi(Z_\pi)$ and $\chi(Z_{\pi/2})$ matrices are presented in
Figs.~\ref{fig_n04}(a)-\ref{fig_n04}(d), also agreeing excellently with the results in an ideal scenario.
The process fidelities of these two STA  gates are $F_P(Z_\pi)=95.23\%$
and $F_P(Z_{\pi/2})=95.20\%$. After excluding errors in state preparation and readout,
the gates fidelities are $F_g(Z_\pi)=99.89\%$ and $F_g(Z_{\pi/2})=99.87\%$.

\begin{figure}[tp]
\includegraphics[width=0.65\columnwidth]{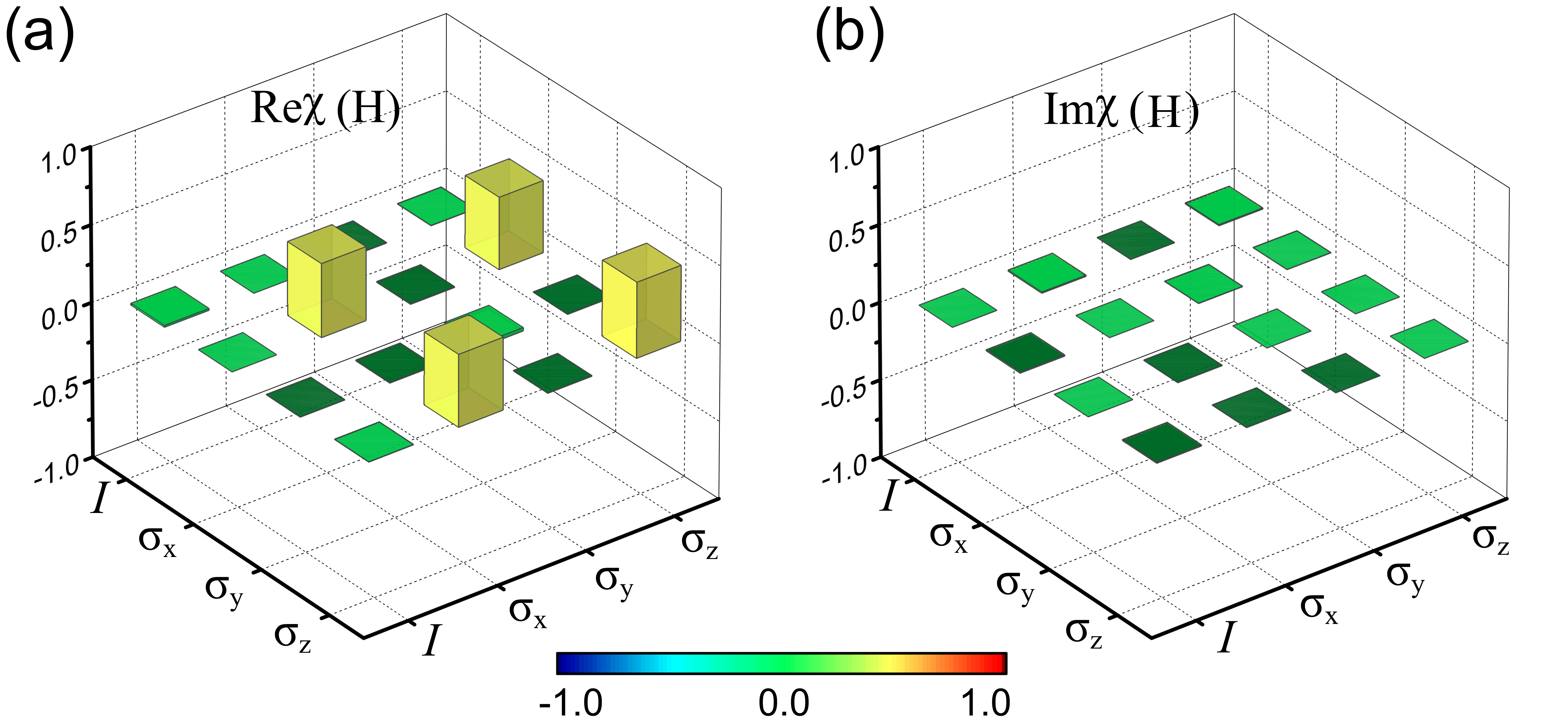}
\caption{The experimental measurement of the $\chi$ matrix for the Hadamard gate: (a) real and (b) imaginary parts.}
\label{fig_n05}
\end{figure}

\subsection{Hadamard Gate}
\label{sec4c}

An arbitrary single-qubit quantum gate can be realized by a combination of sequential rotations about $X$-, $Y$- and $Z$-axes.
For example, the Hadamard gate can be generated by $\pi/2$ rotation about the $Y$-axis followed by $\pi$ rotation about the $X$-axis~\cite{ChuangBook}, i.e.,
\be
U_H= U_{X_\pi} U_{Y_{\pi/2}} = \frac{\sqrt{2}}{2} \left( \ba{cc} 1 & 1\\
                                       1 & -1 \ea \right)
\label{eq_24}
\ee
In the STA protocol, the Hadamard gate can be realized by a one-step operation, which reduces the errors accumulated through multiple steps.
Our reference `adiabatic' field $\bm B_0(t)$ is designed as
\be
\left\{\ba{ccl} B_{0; x}(t) &=& \frac{\sqrt{2}}{2}\Omega(t) \cos\varphi(t)  \\
B_{0; y}(t) &=& \Omega(t) \sin\varphi(t)\\
B_{0; z}(t) &=& -\frac{\sqrt{2}}{2} \Omega(t) \cos\varphi(t)
 \ea \right.
\label{eq_25}
\ee
with $\Omega(t) = A \sin(2\pi t/T)$ and $\varphi(t)=(\pi/2) [1-\cos(\pi t/T)]$. After including counter-diabatic field $\bm B_\mathrm{cd}(t)$ and the DRAG field $\bm B_\mathrm{d}(t)$,
we perform the same QPT measurement as above. The experimentally reconstructed $\chi(H)$ matrix is displayed
in Figs.~\ref{fig_n05}(a) and~\ref{fig_n05}(b). The process fidelity is $F_P(H)=94.93\%$
while the gate fidelity with the errors in state preparation and readout excluded is $F_g(H) = 99.81\%$.

\subsection{Interleaved Randomized Benchmarking Measurement}
\label{sec4d}

\begin{figure}[tp]
\includegraphics[width=0.55\columnwidth]{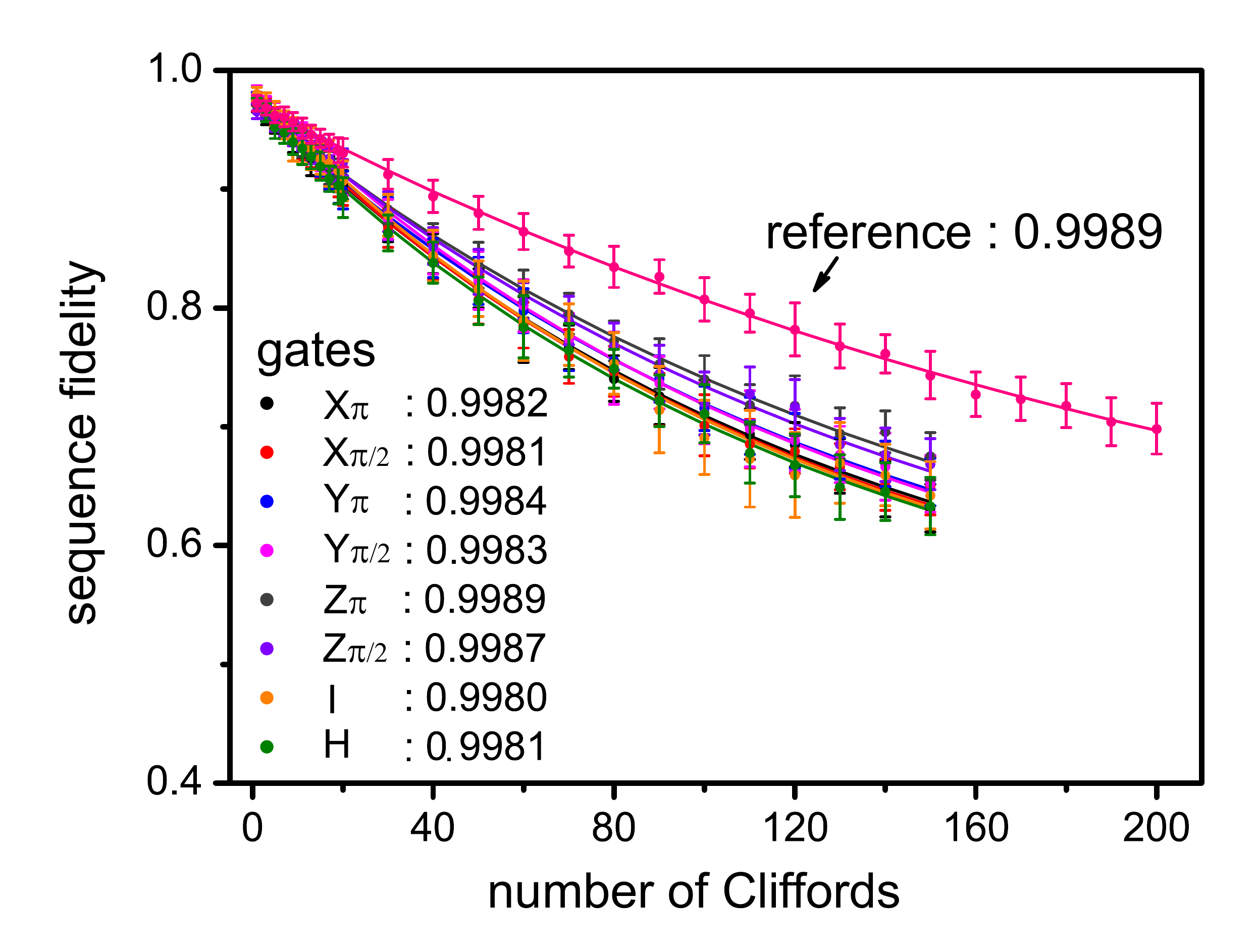}
\caption{Randomized benchmarking measurement for a set of single-qubit STA  quantum gates.
The reference and interleaved sequence fidelities are displayed as a function of the number of Cliffords.
Each sequence fidelity is averaged over $k=50$ randomized operation (see text), with its standard deviations from the mean shown as an error bar. All the gate fidelities are calculated and shown in the figure.
 }
\label{fig_n06}
\end{figure}

In the QPT measurement, the errors of state preparation and readout are mixed with the error of a quantum gate operation.
To extract the gate fidelity, we perform the Clifford-based randomized benchmarking measurement~\cite{KnillPRA08,MagesanPRL12_RB,ChowPRL09,BarendsNat14,SheldonPRA16}.
For a single qubit, the Clifford group consists of 24 rotations preserving the octahedron in the Bloch sphere. 
In principle, each Clifford operator can be realized by a combination from the elements of
$\{I, X_{\pi}, X_{\pm \pi/2}, Y_{\pi}, Y_{\pm \pi/2}\}$. The qubit is initially prepared at the spin-up state ($|\uparrow\rangle=|0\rangle$),
and then driven by a sequence of  $m$ randomly selected Clifford gates. The combined operation is
described by a unitary matrix, $U_C = \prod_{i=1}^m U_i$. Since the Clifford group is a closed set, $U_C$ is always a
Clifford operator. Subsequently, the $(m+1)$-th step is the reversed step of $U_C$ and the total quantum operation
is written as
\be
U_\mathrm{tot} = U^+_C \prod_{i=1}^m U_i.
\label{eq_26}
\ee
The remaining population $P_0(t_f)$ of the initial state is measured afterwards. After repeating the above random operation sequence
$k$ (= 50 in our experiment) times, we calculate the average result of $P_0(t_f)$, which represents a sequence fidelity, $F_\mathrm{seq}(m)$.
As shown in Fig.~\ref{fig_n06}, this sequence fidelity can be well fitted by a power-law decaying function~\cite{MagesanPRL12_RB},
\be
F_\mathrm{seq} = A_0 p^m+B_0,
\label{eq_27}
\ee
where $A_0$ and $B_0$ absorbs the errors in state preparation and readout, and $p$ is a depolarizing parameter.
The average error over the randomized Clifford gates is given by~\cite{MagesanPRL12_RB}
\be
r = \frac{d-1}{d}(1-p)
\label{eq_28}
\ee
where $d=2^N$ is the dimension of the Hilbert space for an array of $N$ qubits. In our experiment, the value of the
average error is $r=0.0011$, or equivalently the fidelity of a randomized Clifford gate is 99.89\%,
which serves as a reference for our next interleaved operation (see Fig.~\ref{fig_n06}).

To extract the fidelity of a specific gate $U_g$, we make an interleaved operation~\cite{MagesanPRL12_RB}. At each step, the qubit is driven
by a combination of a randomly select Clifford operator followed $U_g$.
With the product operator, $U^\pr_C = \prod_{i=1}^m (U_g U_i)$, and the $(m+1)$-th operator of $(U^\pr_C)^+$, the
total quantum operation is described by $U^\pr_\mathrm{tot}=(U^\pr_C)^+\prod_{i=1}^m (U_g U_i)$~\cite{BarendsNat14,MagesanPRL12_RB}.
Similarly, we measure the sequence fidelity $F^\pr_\mathrm{seq}(m)$. As shown by the examples in Fig.~\ref{fig_n06}, $F^\pr_\mathrm{seq}(m)$
can also be well fitted by Eq.~(\ref{eq_27}) with a new depolarizing parameter $p^\pr$.
Here $p^\pr$ can be considered as a product of the average number $p$ of a
randomized Clifford operator and the intrinsic number $p_g$ of the specific gate $U_g$, i.e., $p_g = p^\pr/p$.
Substituting $p_g$ into Eq.~(\ref{eq_28}), we obtain the intrinsic error $r_g$ and the gate fidelity of $U_g$
is given by
\be
F_g  = 1-\frac{d-1}{d}\left(1-\frac{p^\pr}{p}\right).
\label{eq_29}
\ee
In Fig.~\ref{fig_n06}, we list the results of 8 example STA gates, and all the values of $F_g$ are equal or
greater than 99.8\%. Notice that the fidelity of the Harmard gate ($F_g(H)= 99.81\%$) is higher than the product
of the fidelities of the $Y_{\pi/2}$ and $X_\pi$ gates ($F_g(X_\pi)F_g(Y_{\pi/2})=99.65\%$). Thus, our one-step
STA gate can efficiently reduce the error accumulation in a combined operation of multiple gates.

\section{Summary}
\label{sec5}

In this paper, we propose a scheme of building a universal quantum gate using 
a `shortcut-to-adiabaticity' trajectory, which shares the same spirit as
in Ref.~\cite{LiangPRA16} but with a different design.
This scheme is successfully implemented in a high-quality superconducting Xmon qubit,
and various single-qubit STA quantum gates are created through a precise microwave control.
As demonstrated by the examples of rotations about $X$- and $Z$-axes and the Hadamard gate,
we have achieved high process and gate fidelities ($F_p> 94.9\%$ and $F_g\ge 99.8\%$),
which are very close to the state-of-the-art values ($F_g\ge 99.91\%$) in the superconducting
Xmon qubit system. In principle, the STA quantum gates allow a large flexibility in
the control parameters, such as the pulse amplitude, operation time and pulse shape.
Although this paper is focused on single-qubit gates, the STA scheme can be extended
to a multi-qubit system~\cite{BarendsNat14,LiangPRA16}. The improvement and extension of our STA quantum gates
will be addressed in the near future.

\begin{acknowledgements}

The work reported here is supported by the National Basic Research Program of China (2014CB921203, 2015CB921004),
the National Key Research and Development Program of China (2016YFA0301700, 2017YFA0304303),
the National Natural Science Foundation of China (NSFC-11374260, 21573195, 11625419, 11474177),
the Fundamental Research Funds for the Central Universities in China, and the Anhui Initiative in Quantum Information Technologies (AHY080000).
This work was partially carried out at the University of Science and Technology of China Center for Micro and Nanoscale Research and Fabrication.

\end{acknowledgements}


\begin{thebibliography}{40}

\bibitem{ChuangBook}
Nielsen M A and Chuang I L 2000 {\it Quantum computation and quantum information} (Cambridge: Cambridge University Press)


\bibitem{LaddNat10}
Ladd T D, Jelezko F, Laflamme R, Nakamura Y, Monroe C and O'Brien J L 2010 {\it Nature} {\bf 464} 45-53

\bibitem{RyanNJP09}
Ryan C, Laforest M and Laflamme R 2009 {\it New Journal of Physics} {\bf 11} 013034

\bibitem{LuNPJQI17}
Lu D et al 2017 {\it npj Quantum Information} {\bf 3} 1

\bibitem{BenhelmNatPhys08}
Benhelm J, Kirchmair G, Roos C F and Blatt R 2008 {\it Nat. Phys.} {\bf 4} 463-66

\bibitem{BarendsNat14}
Barends R {et~al} 2014 {\it Nature} {\bf 508} 500-3


\bibitem{SjoqvistPhy08}
Sj\"{o}qvist E 2008 {\it Physics} {\bf 1} 35



\bibitem{ZhuPRA05}
Zhu S L and Zanardi P 2005 {\it Phys. Rev. A} {\bf 72} 020301





\bibitem{DemirplakJPCA03}
Demirplak M and Rice S A 2003 {\it J. Phys. Chem. A} {\bf 107} 9937-45

\bibitem{BerryJPhysA09}
Berry M 2009 {\it J. Phys. A} {\bf 42} 365303

\bibitem{XChenPRL2010}
Chen X, Lizuain I, Ruschhaupt A, Gu{\'e}ry-Odelin D and Muga J 2010 {\it Phys. Rev. Lett.} {\bf 105} 123003

\bibitem{MasudaPRS10}
Masuda S and Nakamura K 2010 {\it Proc. R. Soc. A} {\bf 466} 1135-54

\bibitem{Torrontegui13AAMOPhy}
Torrontegui E, Ib\'{a}\~{n}ez S, Mart\'{i}nez-Garaot S, Modugno M, del Campo A, Gu\'{e}ry-Odelin D, Ruschhaupt A, Chen X and Muga J G
2013 {\it Adv. At. Mol. Opt. Phys.} {\bf 62} 117

\bibitem{CampoPRL12}
del Campo A, Rams M M and Zurek W H 2012 {\it Phys. Rev. Lett.} {\bf 109} 115703


\bibitem{CampoPRL13}
del Campo A 2013 {\it Phys. Rev. Lett.} {\bf 111} 100502

\bibitem{TongSR15}
Zhang J, Kyaw T H, Tong D M, Sj\"{o}qvist E and Kwek L C 2015 {\it Sci. Rep.} {\bf 5} 18414

\bibitem{SantosSciRep15}
Santos A C and Sarandy M S 2015 {\it Sci. Rep.} {\bf 5} 15775




\bibitem{BasonNatPhy11}
Bason M G, Viteau M, Malossi N, Huillery P, Arimondo E, Ciampini D, Fazio R, Giovannetti V, Mannella R and Morsch O
2012 {\it Nat. Phys.} {\bf 8} 147-52


\bibitem{JFZhangPRL13}
Zhang J F {et~al} 2013 {\it Phys. Rev. Lett.} {\bf 110} 240501

\bibitem{AnNatCommu16}
An S M, Lv D S, del Campo A and Kim K 2016 {\it Nat. Commun.} {\bf 7} 12999

\bibitem{DuYXNatCommu16}
Du Y X, Liang Z T, Li Y C, Yue X X, Lv Q X, Huang W, Chen X, Yan H and Zhu S L 2016 {\it Nat. Commun.} {\bf 7} 12479


\bibitem{ZhouNatPhys16}
Zhou B B, Baksic A, Ribeiro H, Yale C G, Heremans F J, Jerger P C, Auer A, Burkard G, Clerk A A and Awschalom D D
2017 {\it Nat. Phys.} {\bf 13} 330-4


\bibitem{ZZXPRA17}
Zhang Z X, Wang T H, Xiang L, Yao J D, Wu J L and Yin Y 2017 {\it Phys. Rev. A} {\bf 95} 042345

\bibitem{SCPMA2018}
Wang T H, Zhang Z X, Xiang L, Gong Z H, Wu J L and Yin Y 2018 {\it Sci. China Phys. Mech.} {\bf 61} 047411


\bibitem{LiangPRA16}
Liang Z T, Yue X X, Lv Q X, Du Y X, Huang W, Yan H and Zhu S L
2016 {\it Phys. Rev. A} {\bf 93} 040305


\bibitem{MotzoiPRL09}
Motzoi F, Gambetta J M, Rebentrost P and Wilhelm F K 2009 {\it Phys. Rev. Lett.} {\bf 103} 110501


\bibitem{GambettaPRA11}
Gambetta J M, Motzoi F, Merkel S T and Wilhelm F K 2011 {\it Phys. Rev. A} {\bf 83} 012308


\bibitem{LuceroPRA10}
Lucero E {et~al} 2010 {\it Phys. Rev. A} {\bf 82} 042339


\bibitem{ChowPRA10}
Chow J M, DiCarlo L, Gambetta  J M, Motzoi F, Frunzio L, Girvin S M and Schoelkopf R J 2010 {\it Phys. Rev. A} {\bf 82} 040305



\bibitem{BarendsPRL13}
Barends R {et~al} 2013 {\it Phys. Rev. Lett.} {\bf 111} 080502



\bibitem{KellyNat15}
Kelly J {et~al} 2015 {\it Nature} {\bf 519} 66-9



\bibitem{DunsworthAPL17}
Dunsworth A {et~al} 2017 {\it Appl. Phys. Lett.} {\bf 111} 022601


\bibitem{RoyAPL15}
Roy T, Kundu S, Chand M, Vadiraj A M, Ranadive A, Nehra N, Patankar M P, Aumentado J, Clerk A A and Vijay R
2015 {\it Appl. Phys. Lett.} {\bf 107} 262601


\bibitem{XiaoYuanPRL16}
Yuan X, Liu K, Xu Y, Wang W, Ma Y W, Zhang F, Yan Z P, Vijay R, Sun  L Y and Ma X F
2016 {\it Phys. Rev. Lett.} {\bf 117} 010502

\bibitem{JohnsonPRL12}
Johnson J E, Macklin C, Slichter D H, Vijay R, Weingarten E B, Clarke J and Siddiqi I 2012 {\it Phys. Rev. Lett.} {\bf 109} 050506




\bibitem{BialczakNatPhys10}
Bialczak R C {et~al} 2010 {\it Nat. Phys.} {\bf 6} 409-13

\bibitem{YamamotoPRB10}
Yamamoto T {et~al} 2010 {\it Phys. Rev. B} {\bf 82} 184515


\bibitem{KnillPRA08}
Knill E, Leibfried D, Reichle R, Britton J, Blakestad R B, Jost J D, Ozeri R, Seidelin S and Wineland D J 2008 {\it Phys. Rev. A} {\bf 77}, 012307

\bibitem{MagesanPRL12_RB}
Magesan E {et~al} 2012 {\it Phys. Rev. Lett.} {\bf 109} 080505

\bibitem{ChowPRL09}
Chow J M, Gambetta J M, Tornberg L, Koch J, Bishop L S, Houck A A, Johnson B R, Frunzio L, Girvin S M and Schoelkopf R J
2009 {\it Phys. Rev. Lett.} {\bf 102} 090502

\bibitem{SheldonPRA16}
Sheldon S, Bishop L S, Magesan E, Filipp S, Chow  J M and Gambetta J M 2016 {\it Phys. Rev. A} {\bf 93} 012301






\end{thebibliography}
\end{document}